\documentclass[12pt,preprint]{aastex}

\def\ergcm2s{${\rm\,ergs\,cm^{-2}\,s^{-1}}$}
\newcommand{\obj}{IRAS 00521-7054}
\def\Fe{Fe K$\alpha$}

\begin{document}

\title{A possible ultra strong and broad Fe K$\alpha$ emission line in
  Seyfert 2 galaxy IRAS 00521-7054}

\author{Y. Tan, J. X. Wang, X. W. Shu, and Youyuan Zhou}

\affil{CAS Key Laboratory for Research in Galaxies and Cosmology, Department of Astronomy, University of Science and Technology of China, Hefei, Anhui 230026, China;  tanyin29@mail.ustc.edu.cn; jxw@ustc.edu.cn; xwshu@mail.ustc.edu.cn; yyzhou@ustc.edu.cn}

\begin{abstract}
We present XMM-Newton spectra of the Seyfert 2 Galaxy IRAS 00521-7054. 
A strong feature at $\sim6$ keV (observer's frame) can be formally fitted with a strong (EW=$1.3\pm0.3$~keV in the rest frame) and broad Fe K$_{\alpha}$ line, extending down to 3~keV.
The underlying X-ray continuum could be fitted with an absorbed powerlaw (with $\Gamma$ = 1.8$\pm0.2$ and N$_H$ = 5.9$^{+0.6}_{-0.7}$ $\times~10^{22}$ cm$^{-2}$) plus a soft component. 
If due to relativistically smeared
reflection by an X-ray illuminated accretion disk, the spin of
the supermassive black hole is constrained to be
$0.97^{+0.03}_{-0.13}$ (errors at 90\% confidence level for one 
interesting parameter),
and the accretion system is viewed at an inclination angle of  37$\pm4$$^{\circ}$.
This would be the first type 2 AGN reported with strong red \Fe\ wing detected which demands a fast rotating SMBH. The unusually large EW would suggest that the light bending effect is strong in this source. Alternatively, the spectra could be fitted by a dual absorber model (though with a global $\chi^2$ higher by $\sim6$ for 283 d.o.f) with N$_{H1}$ = 7.0$\pm0.8$ $\times$ $10^{22}$ cm$^{-2}$ covering 100\% of the X-ray source, and N$_{H2}$ =  21.7$^{+5.6}_{-5.4} $$\times~10^{22}$ cm$^{-2}$ covering 71\%, 
which does not require an extra broad Fe K$_{\alpha}$ line. 

\end{abstract}

\keywords{galaxies: active --- galaxies: Seyfert --- X-rays: individual (IRAS 00521-7054)}

\section{Introduction}
The broad Fe K$\alpha$ fluorescent emission line in Active Galactic Nuclei (AGNs), first clearly detected by ASCA \citep{tanaka95}, provides an important tracer to probe the physical conditions and geometrical distribution of matter in the vicinity of supermassive black hole. The line is believed  from the accretion disk by reflecting the illuminating X-ray continuum, and the unique broad and skewed (toward the red) line profile is a consequence of the relativistic Doppler effect and strong gravitational redshift
\citep[e.g.][]{fabian89,laor91}.  Such broad Fe K$\alpha$ emission lines have been detected in a significant number of AGNs \citep{nandra1997,wang99,fabian2002,
nandra2007,miller07,guainazzi2010a,de10}.
More interestingly, in a few sources the red wing of the Fe K$\alpha$ emission line appears too broad to be produced from accretion disk surrounding Schwarzschild black holes \citep{wilms01,fabian2002,patrick2011}. In these cases, assuming the accretion disk extends to the inner-most stable circular orbit (ISCO, determined by the spin of black hole), fast rotating  Kerr black holes are required, for which the accretion disk could
extend closer to the horizon thus stronger gravitational redshift could be produced
\citep{reynolds2008}. Therefore, the Fe K$\alpha$ line profile enables astronomers to measure the spin of SMBHs, which is otherwise extremely difficult or even impossible \citep{reynolds2003}. Till now, rotating SMBHs have been claimed in a number of type 1 AGNs  \citep[e.g.][]{brenneman06,brenneman2009,patrick2011,brenneman2011}. However, strong debates exist on the nature of the ultra broad red wings, challenging the base of such measurements. An alternative scenario is that the observed broad and skewed line profile is
artificial due to continuum curvature  produced  by more complex absorption \citep[e.g. partial covering ionized absorber,][]{reeves04,miller2008,miller2009,turner09}.
Although various efforts have been made to distinguish two scenarios, including performing spectral fitting over wider band to better determine the absorption and the underlying continuum \citep[]{patrick2011b}, studying bare Seyferts with no or very weak warm absorption \citep[]{patrick2011}, and monitoring the lag between reflection component and the continuum \citep[]{fabian2009},  a clear picture has not been obtained.

Meanwhile, type 2 AGNs could provide a different insight into the nature of broad \Fe\ emission.
According to unification model, type 2 AGNs are viewed at higher inclination angle comparing with type 1 sources. Furthermore, while
ionized absorption, which plays the key role in the alternative scenario to the broad \Fe\ line, is common in type 1 AGNs \citep[e.g.][]
{reynolds1997}, X-ray absorption in type 2 sources is dominated by cold matter.
Till now, broad \Fe\ lines were detected only in a few type 2 AGNs \citep[see Guainazzi et al. 2010 for a summary of broad \Fe\ line detections in obscured AGNs, mostly type 1.8 and 1.9 sources;][]{iwasawa1996MNRAS.279..837I,dawangan2003,shu10a}.
However, neither of the \Fe\ line red wing in these type 2 AGNs is broad enough to demand a
rotating SMBH.

\obj\ is a Seyfert 2 galaxy at $z$ = 0.0689. In this letter we present new XMM-Newton observation on \obj\ taken in 2006, in which we detect an ultra strong and broad \Fe\ line. 
In \S 2 we describe the
XMM observations and data reduction. \S3 provides our detailed spectral fitting to the continuum and the emission line,
followed by discussion in \S4.

\section{Data Reduction}

\obj\ was observed with XMM-Newton twice in 2006 in Prime Full Window mode. The durations of the two exposures were 17.3 ks (on March 22, 2006) and 14.2 ks (April 22, 2006) respectively. We reduced the raw data with the XMM-Newton Scientific Analysis System (SAS) \citep{gabriel04} version 10.0.0. The exposures were processed using the SAS pipeline ``epchain" and  ``emchian". We filtered the event files to include only those events with XMM-SAS quality flag FLAG==0 ($\#$XMMEA\_EM) for PN (MOS1 and MOS2),
and extracted single- and double-pixel events (PATTERN $\le$ 4) for PN, and single- to quadruple-pixel events (PATTERN $\le$ 12) for MOS1 and MOS2.
After excluding high background intervals,
the net exposures for two observations  are 5.7 ks and 7.9 ks for PN, and 7.4 ks and 11.0 ks for MOS respectively.
Source spectra were extracted from circles with radius of 30'' for PN (to avoid nearby chip edge) and 40'' for MOS   
 centered on the source position. For both PN and MOS data, 
the background events were extracted from source-free areas on the same chip using four circular 
circles with a combined area 4 times larger than the source region. 
We note that adopting different background regions would not alter any scientific results presented in this work.
The response matrix and ancillary response file were generated using the RMFGEN and ARFGEN tools with the SAS software.
The extracted net source count rate is $\sim$ 0.26 cts/s for PN, and $\sim$ 0.08 cts/s for MOS, thus the exposures are free from pile-up effect.

\section{Spectral Fitting}

We examined the light curves of each exposure, and found no significant variation within or between 
two exposures. Through a preliminary spectral fitting, we also found no spectral variations between two
observations.
We therefore 
combined the spectra from two exposures to increase the S/N ratio in the spectra, and obtained one composite PN spectrum and one MOS spectrum (co-added from MOS1 and MOS2)\footnote{ Fitting spectra from two exposures simultaneously yields results consistent with what we presented in this paper, but is rather computer time consuming while adopting complex models.}. Both spectra were binned to have at least 20 counts per bin to allow the use of $\chi^2$ statistics with XSPEC V12.6.0.
All fitting errors are of $90\%$ confidence level for one parameter ($\Delta{\chi}^2=2.71$), if not otherwise stated.
The adopted cosmological parameters are $H_0=70 \ km\ s^{-1} Mpc^{-1}$, $\Omega_\Lambda=0.73$ and $\Omega_m=0.27$.
During spectral fitting, the Galactic column density along the line of sight to {\obj}
is always taken into account \citep[$N_{\rm H}=4.44 \times10^{20}\ cm^{-2}$, ][]{dickey90}.

In Figure 1 we present XMM spectra of \obj\ { for both PN and the summed MOS1 and MOS2 data. 
It can be seen that the MOS data are consistent with the PN data within the statistical errors.}
We first fit the 0.5 -- 10 keV spectra with an absorbed powerlaw plus an extra soft component. The soft component could be scattered emission of the intrinsic X-ray continuum, or from the host galaxy. We find that either a powerlaw free from intrinsic absorption (with $\Gamma$ = 2.4$\pm0.5$), a blackbody (with T = 0.21$\pm$0.03 keV) or a bremss plasma emission could provide adequate fit to the soft component, and these models are statistically indistinguishable based on current data.
Hereafter, we simply choose a blackbody model to fit the soft component.
We note that choosing different soft component models would not alter the fitting results in this work.
{ This simple continuum model provides a poor fit to the data, with $\chi^2=$368.3 for 
287 degrees of freedom.  The data to model ratio plot (Figure 1b) shows that 
there is  clear residual emission between 3 -- 7 keV that could be interpreted as the existence of a 
relativistic \Fe\ emission line}. We then exclude data bins
between 3 -- 7 keV to fit the continuum, and present the plot of data to best-fit model ratio in Fig. 1 (Panel c) to demonstrate the line profile.

The line appears rather broad, and the red wing extends to as low as 3 keV.
We then add a relativistic disk line model  \citep[$laor$ in XSPEC,][]{laor91} to fit the broad line (see Fig. 1), and the fitting is significantly improved ($\Delta$$\chi^2$ =  58 for 4 more free parameters).  The best-fit line and continuum parameters are listed in Table 1. During the fit, we fixed the line energy at 6.4 keV in the source rest frame, since it is otherwise poorly constrained by the data. The small best-fit inner radius
(R$_{in}$ = 1.8$^{+0.7}_{-0.8}$ r$_g$) suggests that the central SMBH is
fast rotating (fitting with a $diskline$ model for a non-rotating SMBH instead yields
larger $\chi^2$, $\Delta\chi^2$ = 9). 
Furthermore, the line EW is remarkably large (EW = 1.3$\pm0.3$ keV in the rest frame). The best-fit inclination of the disk is 37$^{+6}_{-5}$. Since narrow \Fe\ cores are often seen
in the X-ray spectra of AGNs (Yaqoob \& Padmanabhan 2004; Shu et al. 2010b, 2011), we added a narrow Gaussian line (with central energy fixed at 6.4 keV in the rest frame, and line width at zero, i.e., unresolved by XMM PN and MOS spectra) to the fit, and found such a narrow component is statistically not required in additional to the broad component ($\Delta$$\chi^2$ $<$ 1, EW $<$ 56.0 eV). Reflection component from neutral material in additional to the underlying powerlaw \citep[$pexrav$ in XSPEC,][]{magdziarz95} is statistically not required either ($\Delta$$\chi^2$ = 0, R $<$ 0.8),  the normalization of which thus has little effect on the broad Fe Ka line intensity.

We also apply a self-consistent model \citep[$kerrconv$ $\times$ $reflionx$, ][]{brenneman06, ross05} to fit the broad line (see Fig. 2), and the best-fit parameters are presented in Table 1.
The fitting also suggests that the central SMBH is rotating, with $\alpha$ = 0.97$^{+0.03}_{-0.13}$. This model produces similar residual $\chi^2$/dof comparing with the simple
$laor$ line model (310.13/283 vs 311.54/280), and consistent physical parameters,
including inclination angle $\theta$, inner radius and emissivity index $q$.

\section{Discussion}

We present the detection of an ultra strong and broad \Fe\ line in Seyfert 2 galaxy \obj.
The X-ray continuum is fitted with a powerlaw plus a soft blackbody emission. The observed 0.5 -- 2 keV  and 2 -- 10 keV fluxes  are 3.8 $\times~10^{-14}$ ergs cm$^{-2}$ s$^{-1}$ and 2.4 $\times~10^{-12}$ ergs cm$^{-2}$ s$^{-1}$  respectively.
Absorption corrected 2 -- 10 keV rest frame luminosity is 3.9 $\times~10^{43}$ ergs s$^{-1}$.
The \Fe\ line is rather strong with EW =1.3$\pm0.3$ keV, larger than most broad \Fe\ lines
detected in other AGNs (with typical EW of several 100 eV). Spectral fitting suggests that
the SMBH in \obj\ in rotating with a spin of $\alpha$ = 0.97$^{+0.03}_{-0.13}$. This is the first measurement of SMBH spin in a type 2 AGN.

\subsection{Alternative models}

In this sub-section we investigate whether such an ultra strong and broad \Fe\ line
could be in fact due to curvature in underlying continuum caused by more complex absorption. Below we applied different absorption models to the spectra without including an additional broad emission line component.

Single warm absorber ($absori$ in XSPEC), a partially covering warm absorber, or
a partial covering cold absorber model can be easily ruled out because of the much larger $\chi^2$ reduced ($\Delta\chi^2$ $>$ 50, comparing with the $laor$ line model).

We then adopted a dual cold absorber model (with N$_{H1}$ covering 100\%
of the central X-ray source, and N$_{H2}$ partially covering the continuum).
Interestingly, the dual-absorber model fits the spectra much better
($\chi^2$/dof = 319.09$/$285), however it is still statistically worse than the $laor$ line model ($\Delta\chi^2$ = 9).
In Figure 3(b),  we can still see clear residuals at $\sim$ 6 keV in the data to model ratio plot, comparing with the the residuals of best-fit $laor$ and $reflionx$ model (see Fig. 3d and 3e).
Adding a reflection component ($pexrav$) and/or a narrow Gaussian at 6.4 keV to the dual-absorber model
does not change the produced $\chi^2$. Replacing either one or both of the cold absorbers with
ionized absorption yields consistent results and best-fit ionization parameters at zero, indicating more complex warm absorption models do not improve the fitting.
The best-fit parameters of the dual-absorber model are N$_{H1}$ = 7.0$\pm0.8$ $\times~10^{22}$  cm$^{-2}$
with unit coverage, and N$_{H2}$ =  21.7$^{+5.6}_{-5.4}$ $\times~10^{22}$ cm$^{-2}$ with a covering factor $f$ = 0.71$^{+0.06}_{-0.09}$.
We also note that the best-fit powerlaw slope ($\Gamma$ = 2.8$\pm0.1$) appears considerably steep, greater than that of 99.7\% of AGNs in a large XMM sample \citep{scott2011MNRAS.tmp.1298S}.

We note that in our dual absorber model, all element abundances are fixed to the solar 
values. By adopting the $zvphabs$ model in XSPEC, we tried to fit the spectra by
allowing the Fe and Ni abundance to be free parameters, which could likely produce better fit to the strong edge feature at $\sim$ 7 keV in the spectra. 
We found that an overabundance of Fe/Ni (by a factor of $\sim$ 2.6) could slightly improves the fit ($\Delta \chi^2=2.5$), but is statistically not required, thus the abundance of Fe/Ni can not be well constrained.
This model however yields a flatter powerlaw index ($\Gamma=2.3^{+0.6}_{-0.3}$), consistent with most AGNs. 
Figure 3c shows the data to model ratio plot for this model, in which 
weak residuals at $\sim$ 6 keV is still visible, comparing with the $laor$ and $reflionx$ model. 
This model (by allowing the Fe and Ni abundance to be free parameters)
is still statistically slightly worse than the laor line model (the difference of $\chi^2$ is 6.5 with the 
same degrees of freedom). 
 
We note that, while the
relativistic reflection scenario yields formally a slightly better
fit than the double absorption scenario we have presented in this
paper, only follow-up observations with higher spectral resolution
and/or larger sensitive energy bandpass would allow us to discriminate
between them.

\subsection{The nature of the ultra strong \Fe\ line}

Spectral fitting to the broad \Fe\ line suggests that the accretion system is
viewed at an intermediate inclination angle of 37$\pm4$$^{\circ}$, in agreement
with studies of other type 2 AGNs \citep{guainazzi2010b}. It's interesting to note that
\cite{1987ApJ...313L..53F} reported the detection of broad H$\alpha$ emission line (with full width at zero intensity of $>$ 6000 km/s) in its optical spectrum.
This suggests that \obj\ is  a Seyfert 1.9 galaxy, consistent with the
intermediate inclination angle obtained through X-ray spectral fitting.

The measured broad line shows superb large line EW (1.3$\pm0.3$ keV in the rest frame, fitted
with a $laor$ line), much larger than most of those previously detected in either type 1 or type 2 AGNs. This is however a different situation comparing with large narrow \Fe\ line EW ($>$ 1 keV) reported in Compton-thick AGNs \citep[see][]{levenson2006,liu2010}. In Compton-thick AGNs, the large \Fe\ line EW is due to the strong attenuation to the underlying continuum, and the narrow \Fe\ line, which is produced at a much larger scale comparing with the continuum, suffers no or little obscuration.
In \obj, there exists only moderate obscuration to the continuum. Furthermore, the detected \Fe\ line is ultra broad, and is believed from the innermost region of the accretion disk, thus strong attenuation (if there is any) could not enlarge the line EW.

Theoretical calculations have shown that reflection from a constant density slab which is illuminated with a powerlaw spectrum could produce \Fe\ line with EW no larger
than $\sim$ 800 eV \citep[e.g.][]{2002MNRAS.329L..67B}. One possible explanation to the observed large EW in \obj\ is light bending effect. \cite{fabian2003}  reported a similar large line EW (1.4 keV if fitted with
a $laor$ line) in MCG 6-30-15 during a low-flux interval, and it can be simply attributed to
the strong light bending effect which could (with the X-ray continuum source located
closer to the central SMBH) yield weaker observed flux in continuum, but almost constant reflection component.
We note that the fitted emissivity index $q$ (see table 1) is also consistent with
that reported in MCG 6-30-15 during a low-flux interval (3.3$\pm$0.3).
However, it is still puzzling that the line EW in \obj\ remains stably large during two XMM exposures, 
while such large EW was rather rare in other AGNs and was only detected in low flux state in MCG 6-30-15 \citep{fabian2003}, 
and in 1H0707-495 \citep[]{fabian2009,zoghbi10,de10} and IRAS13224-3809 \citep{ponti10}. 

The $reflionx$ model also yields an over abundance of  iron (Fe/solar = 3.4$\pm$1.3).
We  note that similar over abundance in iron was also reported in other AGNs \citep[e.g.][]{ballantyne2003,risaliti2009,patrick2011}.
The best-fit spin of the SMBH is $\alpha=0.97 
^{+0.03}_{-0.13}$ (90\% confidence level), suggesting that the black hole may 
be maximally spinning. However, a Schwarzschild solution is still
consistent with the data if the 99\% confidence level uncertainties are 
considered (see Figure 4). 
Higher S/N spectra are therefore desired to understand the nature
of the ultra strong and broad \Fe\ line in \obj.

\acknowledgments
The work was supported by Chinese NSF through Grant 10825312.

\begin{figure}
\centering
\includegraphics[angle=-90,scale=0.8]{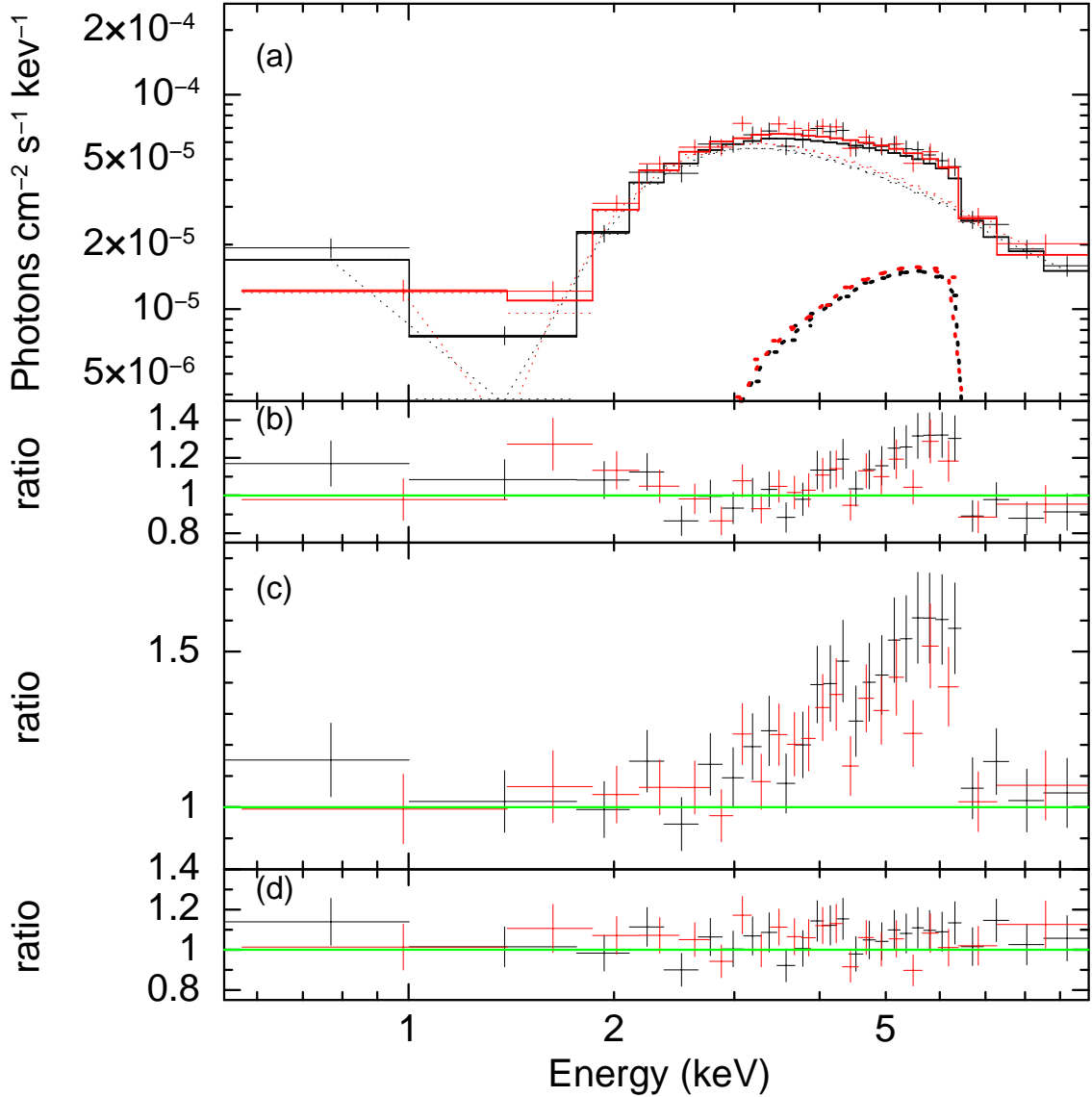}
\caption{ XMM PN (black) and MOS (red) spectra of \obj, with the best-fit 
$laor$ disk line model (Model 1 in table 1) (top panel). 
The spectra were re-binned for display purpose.
(b) Data/model ratio after fitting an absorbed powerlaw plus a soft $bbody$ component to the 0.5 -- 10 keV spectra. 
(c) As (b), but data points between 3 -- 7 keV were excluded during continuum fitting to demonstrate the broad Fe K$\alpha$ line profile. 
(d) Residuals of XMM spectra to the best fit $laor$ disk line model.}
\end{figure}

\clearpage

\begin{figure}
\includegraphics[angle=-90,scale=0.8]{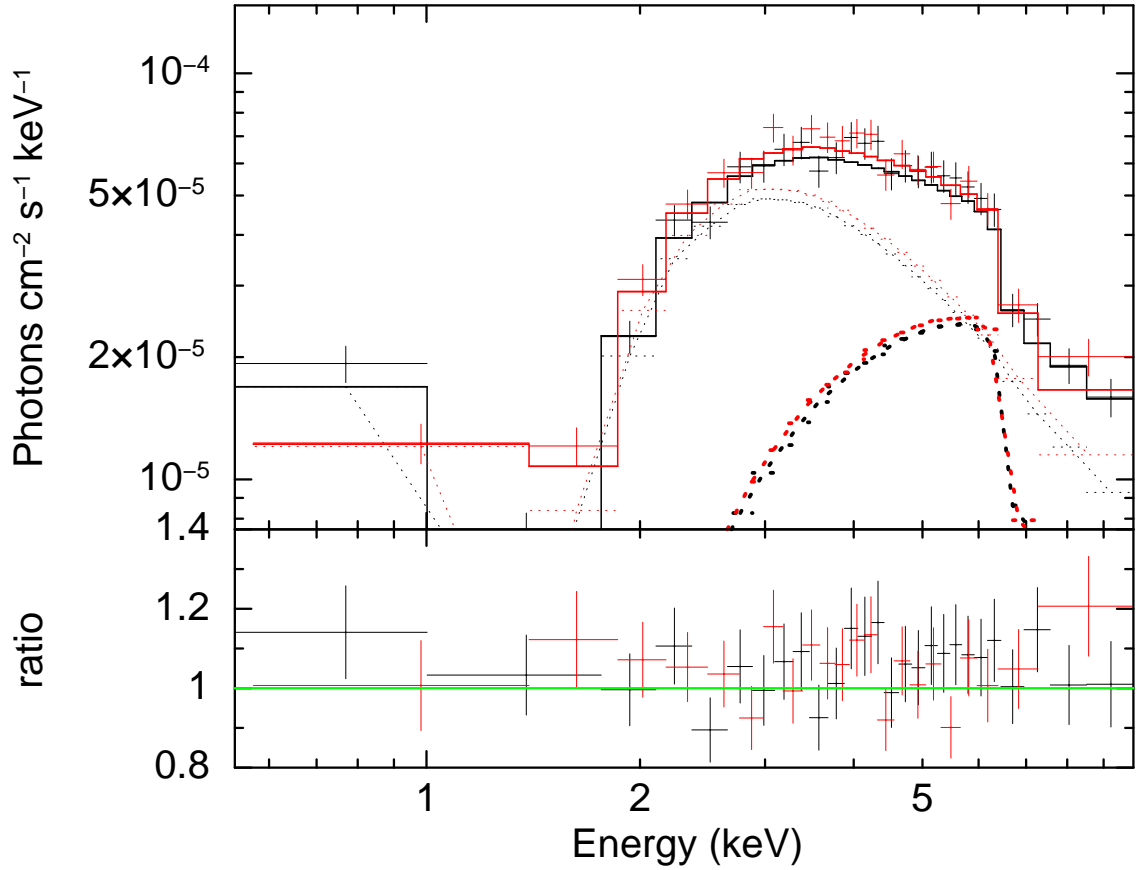}
\caption{XMM spectra of \obj with the best-fit  $reflionx$ model (Model 2 in table 1).}
\end{figure}

\begin{figure}
\includegraphics[angle=-90,scale=0.8]{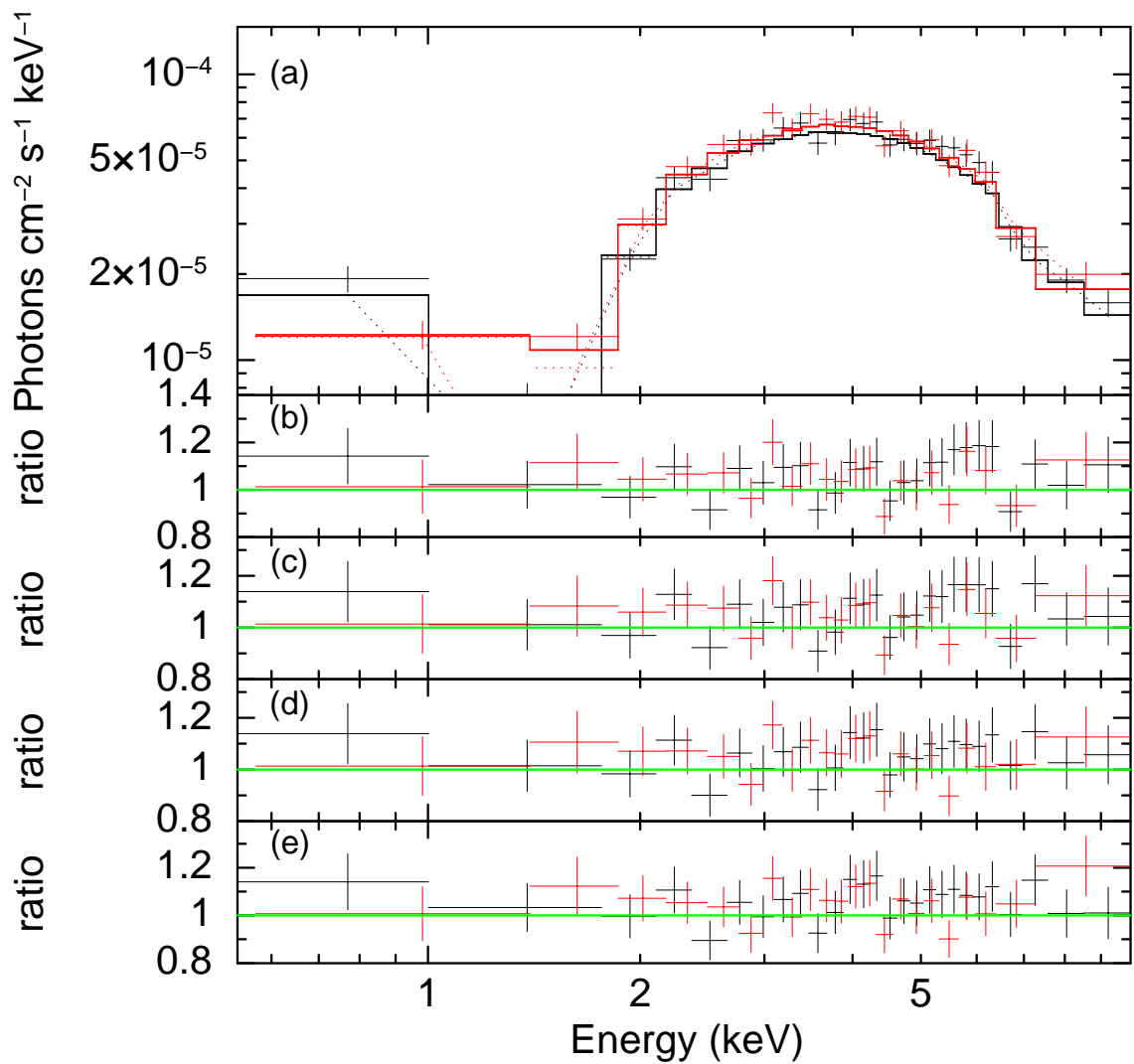}
\caption{XMM spectra of \obj\ with best-fit dual absorber model (Model 3 in table 1) 
(Panel a). (b): The corresponding data/model ratios. (c): the data to model plot for best fit dual absorber model allowing Fe/Ni abundance to be free.
(d) and (e): Residuals after fitting with $laor$ and $reflionx$ model, respectively (see also Figure 1 and 2). 
}
\end{figure}

\begin{figure}
\includegraphics[scale=0.8]{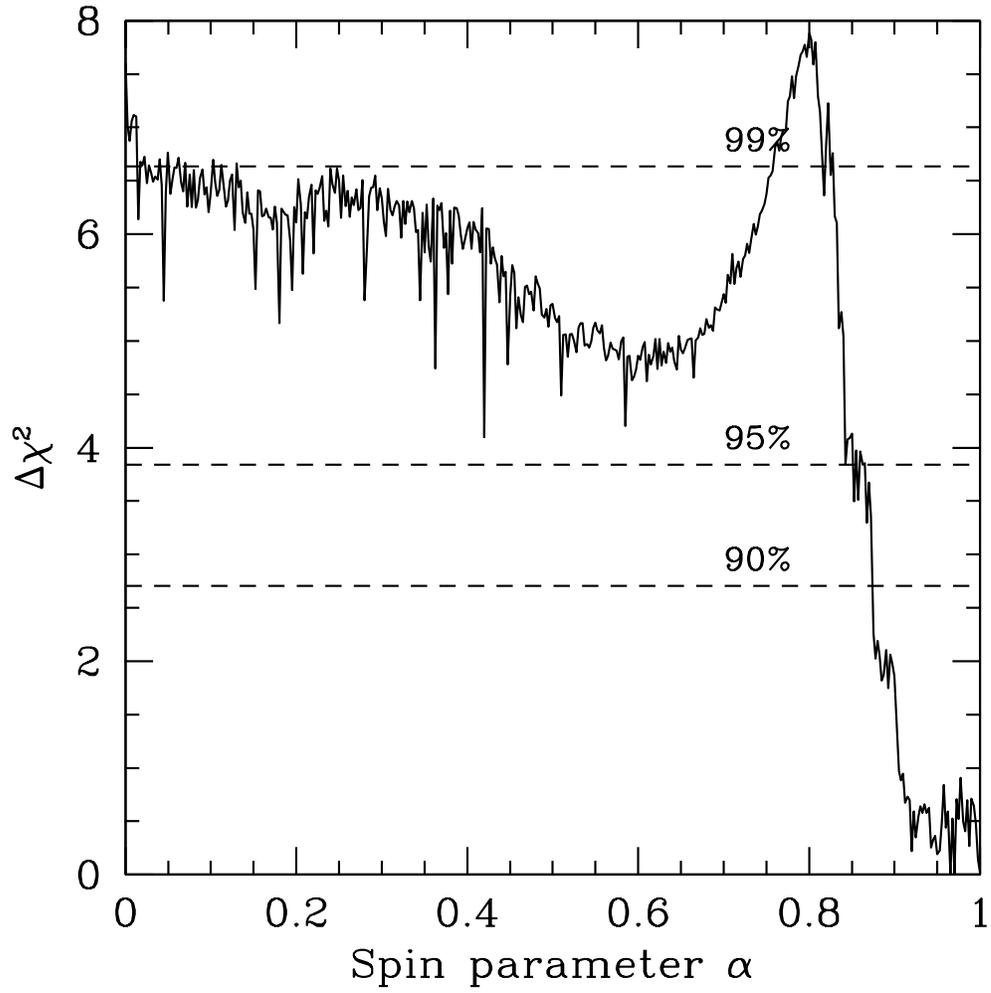}
\caption{$\Delta$ $\chi^{2}$ plot versus the spin parameter $\alpha$ for the best-fit $reflionx$ model.}
\end{figure}

\clearpage

\begin{table*}
\centering
\caption{Best-fit parameters of three spectral models.}
\label{tab:spec_modelFits}
\begin{tabular}{@{}llccc}
\hline
Number & Model-parameters & Model 1 &Model 2 & Model 3 \\
\hline

1 & {\tt zblackbody}: KT (keV)  & 0.21$^{+0.03}_{-0.03}$ & 0.21$^{+0.03}_{-0.02}$ & 0.21$^{+0.03}_{-0.03}$ \\
2 & {\tt zphabs}: Column density, $N_{\rm H}$ (cm$^{-2}$)   &  5.9$^{+0.6}_{-0.7}$$\times10^{22}$ & 6.4$^{+0.3}_{-0.2}$$\times10^{22}$ &  7.0$^{+0.8}_{-0.8}$$\times10^{22}$ \\
3 & {\tt zpowerlw}: Photon index, $\Gamma$  &  1.8$^{+0.2}_{-0.2}$ & 2.2$^{+0.1}_{-0.1}$ &  2.8$^{+0.1}_{-0.1}$ \\
4a & {\tt laor}: Line-energy (keV) & $5.99^{fixed}$& &  \\
4b & {\tt laor}: Emissivity index, $q$  & 3.7$^{+0.8}_{-0.4}$ & &  \\
4c & {\tt laor}: Inclination, $\theta$ (degrees)  &  37$^{+6}_{-5}$ & &    \\
4d & {\tt laor}: Inner radius ($r_{\rm g}$)  & 1.8$^{+0.7}_{-0.8}$ & & \\
4e & {\tt laor}: Outer radius ($r_{\rm g}$)  &  $400^{fixed}$ & &  \\
5a & {\tt kerrconv}: Emissivity index, $q$ &  &   3.4$^{+0.4}_{-0.2}$  & \\
5b & {\tt kerrconv}: Inclination, $\theta$ (degrees) & &     37$^{+4}_{-4}$ & \\
5c & {\tt kerrconv}: Spin parameter, $\alpha$ &  & 0.97$^{+0.03}_{-0.13}$ & \\
5d & {\tt kerrconv}: Inner radius, ($r_{\rm ISCO}$) & &  1.00$^{+0.99}_{-0.00}$ & \\
5e & {\tt kerrconv}: Outer radius, ($r_{\rm ISCO}$) & &     $400^{fixed}$ & \\
6a & {\tt reflionx}: Iron abundance & &    3.4$^{+1.3}_{-1.3}$ & \\
6b & {\tt reflionx}: Photon index, $\Gamma$ & &   Tied to param.3 & \\
6c & {\tt reflionx}: Ionization parameter, $\xi$  & &    16.2$^{+3.2}_{-4.0}$ & \\
                    &\hspace{4.6em}(erg cm$^{-2}$ s$^{-1}$) &&   &   \\
7a & {\tt zpcfabs}: Column density, $N_{\rm H}$ (cm$^{-2}$) & & & 21.7$^{+5.6}_{-5.4}$$\times10^{22}$  \\
7b & {\tt zpcfabs}: Covering fraction, f  & & & 0.71$^{+0.06}_{-0.09}$ \\
\hline
$\chi^2/$d.o.f. &    & 310.13$/$283 & 311.54$/$280 & 319.09$/$285 \\
\hline
\end{tabular}
\end{table*}

\clearpage

\begin{thebibliography}{}
\bibitem[Ballantyne et al.(2002)]{2002MNRAS.329L..67B} Ballantyne, D.~R.,Fabian, A.~C., \& Ross, R.~R.\ 2002, \mnras, 329, L67
\bibitem[Ballantyne et al.(2003)]{ballantyne2003} Ballantyne, D.~R.,Vaughan, S., \& Fabian, A.~C.\ 2003, \mnras, 342, 239
\bibitem[Brenneman \& Reynolds(2006)]{brenneman06} Brenneman, L.~W., \& Reynolds, C.~S.\ 2006, \apj, 652, 1028
\bibitem[Brenneman \& Reynolds(2009)]{brenneman2009} Brenneman, L.~W., \& Reynolds, C.~S.\ 2009, \apj, 702, 1367
\bibitem[Brenneman et al.(2011)]{brenneman2011} Brenneman, L.~W.,Reynolds, C.~S., Nowak, M.~A., et al.\ 2011, \apj, 736, 103
\bibitem[de La Calle P\'erez et al.(2010)]{de10}de La Calle P\'erez, I. et al. 2010, A\&A, 524, 50
\bibitem[Dewangan et al.(2003)]{dawangan2003} Dewangan, G.~C.,Griffiths, R.~E., \& Schurch, N.~J.\ 2003, \apj, 592, 52
\bibitem[Dickey \& Lockman(1990)]{dickey90} Dickey, J.~M., \& Lockman, F.~J.\ 1990, \araa, 28, 215
\bibitem[Frogel \& Elias(1987)]{1987ApJ...313L..53F} Frogel, J.~A., \& Elias, J.~H.\ 1987, \apjl, 313, L53
\bibitem[Fabian et al.(1989)]{fabian89} Fabian, A.~C., Rees, M.~J., Stella, L., \& White, N.~E.\ 1989, \mnras, 238, 729
\bibitem[Fabian et al.(2002)]{fabian2002} Fabian, A.~C., Vaughan,S., Nandra, K., et al.\ 2002, \mnras, 335, L1
\bibitem[Fabian \& Vaughan(2003)]{fabian2003} Fabian, A.~C., \& Vaughan, S.\ 2003, \mnras, 340, L28
\bibitem[Fabian et al.(2009)]{fabian2009} Fabian, A.~C. et al.\ 2009, Nature, 459, 540
\bibitem[{{Gabriel} {et~al.} (2004)}]{gabriel04} Gabriel, C., Denby, M., Fyfe, D.~J., et al.\ 2004, Astronomical Data Analysis Software and Systems (ADASS) XIII, 314, 759
\bibitem[Guainazzi(2010)]{guainazzi2010a} Guainazzi, M.\ 2010,\memsai, 81, 226
\bibitem[Guainazzi et al.(2010)]{guainazzi2010b} Guainazzi, M.,Bianchi, S., Matt, G., et al.\ 2010, \mnras, 406, 2013
\bibitem[Iwasawa et al.(1996)]{iwasawa1996MNRAS.279..837I} Iwasawa, K., Fabian,A.~C., Mushotzky, R.~F., et al.\ 1996, \mnras, 279, 837
\bibitem[Laor(1991)]{laor91} Laor, A.\ 1991, \apj, 376, 90
\bibitem[Levenson et al.(2006)]{levenson2006} Levenson, N.~A., Heckman, T.~M., Krolik, J.~H., Weaver, K.~A., {\&Zdot}ycki, P.~T.\ 2006, \apj, 648,111
\bibitem[Liu \& Wang(2010)]{liu2010} Liu, T., \& Wang, J.-X.\ 2010, \apj, 725, 2381
\bibitem[{{Magdziarz} \& {Zdziarski}(1995)}]{magdziarz95} Magdziarz, P., \& Zdziarski, A.~A.\ 1995, \mnras, 273, 837
\bibitem[Miller(2007)]{miller07} Miller, J.~M.\ 2007, \araa, 45, 441
\bibitem[Miller et al.(2008)]{miller2008} Miller, L., Turner, T.~J., \& Reeves, J.~N.\ 2008, \aap, 483, 437
\bibitem[Miller et al.(2009)]{miller2009} Miller, L., Turner,T.~J., \& Reeves, J.~N.\ 2009, \mnras, 399, L69
\bibitem[Nandra et al.(1997)]{nandra1997} Nandra, K., George,I.~M., Mushotzky, R.~F., Turner, T.~J., \& Yaqoob, T.\ 1997, \apj, 477, 602
\bibitem[Nandra et al.(2007)]{nandra2007} Nandra, K., O'Neill,P.~M., George, I.~M., \& Reeves, J.~N.\ 2007, \mnras, 382, 194
\bibitem[Ponti et al.(2010)]{ponti10} Ponti, G., Gallo, L.~C., 
Fabian, A.~C., et al.\ 2010, \mnras, 406, 2591 
\bibitem[{{Reeves} {et~al.}(2004){Reeves}, {Nandra}, {George}, {Pounds}, {Turner} \& {Yaqoob}}]{reeves04} {Reeves} J.~N., {Nandra} K., {George} I.~M., {Pounds} K.~A., {Turner} T.~J., {Yaqoob} T., 2004, \apj, 602, 648
\bibitem[Reynolds et al.(2005)]{reynolds05} Reynolds, C.~S., Young, A.~J., Fabian, A.~C., \& Con-X AGN Science Team 2005, Bulletin of the American Astronomical Society, 37, 1178
\bibitem[Reynolds(1997)]{reynolds1997} Reynolds, C.~S. 1997, MNRAS, 286, 513
\bibitem[Reynolds \& Nowak(2003)]{reynolds2003} Reynolds, C.~S., \& Nowak, M.~A.\ 2003, \physrep, 377, 389
\bibitem[Reynolds \& Fabian(2008)]{reynolds2008} Reynolds, C.~S., \& Fabian, A.~C.\ 2008, \apj, 675, 1048
\bibitem[Ross \& Fabian(2005)]{ross05} Ross, R.~R., \& Fabian, A.~C.\ 2005, \mnras, 358, 211
\bibitem[Risaliti et al.(2009)]{risaliti2009} Risaliti, G.,Miniutti, G., Elvis, M., et al.\ 2009, \apj, 696, 160
\bibitem[Patrick et al.(2011a)]{patrick2011} Patrick, A.~R., Reeves,J.~N., Porquet, D., et al.\ 2011, \mnras, 411, 2353
\bibitem[Patrick et al.(2011b)]{patrick2011b} Patrick, A.~R., Reeves,J.~N., Lobban, A.~P., et al.\ 2011, \mnras, 416, 2725
\bibitem[Scott et al.(2011)]{scott2011MNRAS.tmp.1298S} Scott, A.~E., Stewart,G.~C., Mateos, S., et al.\ 2011, \mnras, 1298
\bibitem[Shu et al.(2010a)]{shu10a} Shu, X.~W., Yaqoob, T.,Murphy, K.~D., et al.\ 2010a, \apj, 713, 1256
\bibitem[Shu, Yaqoob \ & Wang(2010b)]{shu10b} Shu, X.~W., Yaqoob, T.,\& Wang, J.~X.\ 2010b, \apjs, 187, 581
\bibitem[Shu, Yaqoob \& Wang(2011)]{shu11} Shu, X.~W., Yaqoob, T.,\& Wang, J.~X.\ 2011, \apj, 738, 147
\bibitem[{{Tanaka} {et~al.}(1995)}]{tanaka95} Tanaka, Y., Nandra, K., Fabian, A.~C., et al.\ 1995, \nat, 375, 659
\bibitem[Turner \& Miller(2009)]{turner09} Turner, T.~J., \& Miller, L.\ 2009, \aapr, 17, 47
\bibitem[Wang, Zhou \& Wang(1999)]{wang99} Wang, J.~X., Zhou, Y.~Y., \& Wang, T.~G., 1999, \apj, 523, L129
\bibitem[Wilms et al.(2001)]{wilms01} Wilms, J., Reynolds, C.~S., Begelman, M.~C., et al.\ 2001, \mnras, 328, L27
\bibitem[Yaqoob \& Padmanabhan(2004)]{yaqoob2004} Yaqoob, T. \& Padmanabhan, U. 2004, \apj, 604, 63
\bibitem[Zoghbi et al.(2010)]{zoghbi10} Zoghbi, A., Fabian, 
A.~C., Uttley, P., et al.\ 2010, \mnras, 401, 2419 
\end{thebibliography}
\end{document}